%% file: main.tex
\documentclass[sigplan,nonacm,screen]{acmart}
\acmSubmissionID{xx}
\microtypesetup{expansion=false}

\AtBeginDocument{
  }

\usepackage{amsmath}
\usepackage[normalem]{ulem}
\usepackage{subcaption}
\usepackage{graphicx}
\setkeys{Gin}{draft=false}
\graphicspath{{pdfs/}}
\usepackage{filecontents}
\usepackage{enumitem}
\usepackage{pifont}
\usepackage{xcolor}
\usepackage{booktabs}

\usepackage{cleveref}
\crefname{section}{§}{§§}
\Crefname{section}{§}{§§}
\usepackage{bm}
\usepackage{array}
\usepackage{wrapfig}
\usepackage{balance}
\usepackage{multirow}
\usepackage{caption}
\usepackage{threeparttable}
\usepackage{todonotes}
\usepackage{diagbox}
\usepackage{wrapfig}
\usepackage[most]{tcolorbox}    
\usepackage{enumitem}
\usepackage{xspace}
\usepackage{makecell}
\newcommand{\cmark}{\textcolor{green}{\ding{51}}}
\newcommand{\xmark}{\textcolor{red}{\ding{55}}}

\renewcommand{\emph}[1]{\textit{#1}}
\newcommand{\sys}{FASER\xspace}
\newcolumntype{C}[1]{>{\centering\arraybackslash}m{#1}}

\begin{document}
\sloppypar

\title{
\sys: Fine-Grained Phase Management for Speculative Decoding in Dynamic LLM Serving
}

\author{Wenyan Chen}
\affiliation{
  \institution{NTU Singapore}
  \country{Singapore}
}

\author{Chengzhi Lu}
\affiliation{
  \institution{NTU Singapore}
  \country{Singapore}
}

\author{Yanying Lin}
\affiliation{
  \institution{Shenzhen Institutes of Advanced \\
Technology, CAS; UCAS}
  \country{China}
}

\author{Dmitrii Ustiugov}
\affiliation{
  \institution{NTU Singapore}
  \country{Singapore}
}

\begin{abstract}
Speculative decoding (SD) is a widely used approach for accelerating decode-heavy LLM inference workloads. While online inference workloads are highly dynamic, existing SD systems are rigid and take a coarse-grained approach to SD management. They typically set the speculative token length for an entire batch and serialize the execution of the draft and verification phases. Consequently, these systems fall short at adapting to volatile online inference traffic. Under low load, they exhibit prolonged latency because the draft phase blocks the verification phase for the entire batch, leaving GPU computing resources underutilized. Conversely, under high load, they waste computation on rejected tokens during the verification phase, overloading GPU resources.

We introduce \emph{\sys}, a novel system that features fine-grained SD phase management. First, \sys minimizes computational waste by dynamically adjusting the speculative length for each request within a continuous batch and by performing early pruning of rejected tokens inside the verification phase. Second, \sys breaks the verification phase into frontiers, or chunks, to overlap them with the draft phase. This overlap is achieved via fine-grained spatial multiplexing with minimal resource interference. Our \sys prototype in vLLM improves throughput by up to 53\% and reduces latency by up to 1.92$\times$ compared to state-of-the-art systems.
\end{abstract}

\maketitle

\input{sections/1-intro}
\input{sections/2-background}
\input{sections/3-design}
\input{sections/4-implementation}

\input{sections/5-methodology}

\input{sections/6-evaluation}
\input{sections/7-related_work}
\input{sections/8-conclusion}

\balance
\bibliographystyle{ACM-Reference-Format}
\bibliography{refs}

\end{document}

%% file: sections/1-intro.tex
\section{Introduction}
\label{sec:introduction}

Large Language Model (LLM) inference is fundamentally constrained by autoregressive decoding: each token depends on all previously generated tokens, which forces sequential execution and limits parallelism. As a result, token generation remains a key obstacle to efficient LLM serving. Speculative Decoding (SD)~\cite{chen2023accelerating,leviathan2023fast} has been proposed to address this bottleneck by decoupling token generation into two stages: a lightweight draft model generates multiple candidate tokens sequentially, and a larger target model verifies these candidates in a single parallel pass, reducing the decode latency compared to a single large model. 

While SD can be effective under relatively stable operating conditions, production LLM serving workloads are far more volatile: request arrival rates can fluctuate by up to 35$\times$ between peak and valley periods~\cite{stojkovic2025dynamollm,patel2024splitwise,wang2025burstgpt}. Such workload variation continuously shifts the effective batch size and the system's runtime bottleneck, making a single SD configuration difficult to sustain. In particular, SD does not degrade gracefully as load changes, because its draft generation and target-side verification are still managed at relatively coarse granularity.

This mismatch manifests in three ways. First, under small batches or low load, the draft phase can become the critical path: the system must wait for the draft model to iteratively produce speculative tokens before verification can proceed, leaving GPU resources underutilized and inflating request latency. Second, as the batch grows, the bottleneck shifts to target-side verification. As shown in~\cref{sec:characteristics}, the target model verifies all drafted tokens in a single parallel pass, even though some of them are unlikely to be accepted, wasting substantial computation on the inevitably rejected suffix.
Third, a higher acceptance rate does not necessarily imply lower latency. We find that although a higher acceptance rate reduces the number of SD iterations, it does not eliminate the computational waste incurred by verifying large batches containing many tokens from rejected suffixes. 


Existing SD systems mainly optimize performance through coarse-grained strategies, such as improving draft efficiency, increasing token acceptance, reducing the number of tokens sent for verification, or splitting requests into micro-batches to overlap draft and target execution~\cite{zhang2025draft,liu2024optimizing}. While effective in specific scenarios, these approaches remain insufficient for dynamic serving workloads. Specifically, improving acceptance alone cannot resolve the significant verification overhead and resource waste that arise at large batch sizes. Also, although micro-batch parallelism can hide part of the latency, each request still incurs a full draft-verification cycle, making it difficult to provide consistently low latency when batches are small. Together, these results reveal a common limitation of existing SD systems: they manage draft generation and target verification at coarse granularity, even though the bottleneck between the two shifts significantly with workload dynamics. 

These limitations point to two opportunities for improving SD under dynamic workloads. First, \textit{verification need not process all drafted tokens equally}: tokens in the rejection suffix can be identified early and skipped to reduce wasted target-side computation. Second, \textit{draft generation and verification need not remain as two monolithic stages}: they can be pipelined at finer granularity so that verification starts earlier and overlaps with ongoing drafting. Realizing these opportunities, however, requires overcoming two challenges. The first is how to identify the rejected suffix accurately with sufficiently low overhead. The second is how to maintain an effective balance between the draft and verification phases when workload conditions, batch composition, and acceptance behavior change over time.

To address these challenges, we present \sys, an SD system that manages the draft and verification phases at \emph{fine granularity}.
\sys is built on three key ideas. First, it introduces a lightweight token-wise early-exit mechanism to identify the rejected suffix \emph{during} verification, thereby reducing computational overhead. Second, \sys introduces \emph{Frontier}, a fine-grained execution abstraction that enables verification to begin before the draft phase completes and to proceed concurrently with ongoing drafting, reducing the decode latency under low load. 
Third, \sys couples these mechanisms with an online controller that adapts speculative length and draft-target resource partitioning to the current serving condition. Together, these techniques allow \sys to respond to the shifting bottlenecks of SD.

We implement \sys atop vLLM~\cite{kwon2023efficient} and evaluate it with various models and datasets using production inference traces. Experimental results show that \sys delivers up to 53\% higher throughput and 1.92$\times$ lower latency than state-of-the-art SD systems, with the greatest benefits observed under dynamic, highly variable request patterns. 

%% file: sections/2-background.tex
\section{Background}
\label{sec:background}
In this section, we first introduce SD and describe how it accelerates token generation in LLMs. We then analyze the limitations of existing SD approaches and highlight the key opportunities and challenges in optimizing SD for multi-request serving scenarios.

\subsection{Speculative Decoding Basics}
\label{sec:spec_dec_and_batch}
In conventional autoregressive LLM serving, generating $k$ output tokens requires $k$ decoding iterations because each token depends on all previously generated tokens. This strict sequential dependency makes decoding slow, especially for large models. Speculative decoding (SD)~\cite{chen2023accelerating,leviathan2023fast} mitigates this bottleneck by reducing the number of target-model decoding iterations.

\begin{figure}[t]
  \centering
  \includegraphics[width=\linewidth]{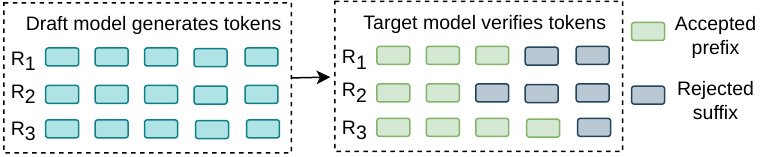}
  \vspace{-1em}
  \caption{Speculative decoding iteration for batched requests, with a speculative token length of 5 for drafting.
  }
  \label{fig:spec_dec_example}
  \vspace{-1.5em}
\end{figure}

The key idea of SD is that verifying multiple candidate tokens in parallel with the target model is often more efficient than generating them strictly one by one. As shown in Fig.~\ref{fig:spec_dec_example}, SD first employs a lightweight \emph{draft} model to produce $k$ candidate tokens conditioned on a prompt prefix. A larger \emph{target} model then verifies these drafted tokens in parallel to determine whether their logits are consistent with the target model's distribution.
Once the target model finishes verifying the entire speculative sequence, the SD system accepts the tokens with matching logits until it encounters the first token with a logit mismatch, discarding the subsequent part of the speculative sequence, which we further refer to as the \emph{rejected suffix}. The system then resamples the first mismatched token using the target model and then resumes the same SD procedure for a new sequence.

The ratio of accepted drafted tokens to the total number of drafted tokens, referred to as the \textit{acceptance rate}, reflects how well the draft model aligns with the target model for each generated token sequence. The decoding process continues with the updated prefix until either the \texttt{EOS} token is generated or the maximum output length is reached. The efficiency of SD, therefore, depends not only on the acceptance rate but also on the costs of draft generation and target-side verification in each iteration.

In production deployments, LLM serving systems deploy instances that process many LLM requests concurrently in continuous batches~\cite{kwon2023efficient}. State-of-the-art SD-enabled  systems~\cite{butler2024pipeinfer,wang2024minions,svirschevski2024specexec,sun2024spectr,xiao2024parallelspec} apply a similar approach: to fully utilize the resources and maximize throughput, the draft model first generates speculative tokens for all requests in a batch simultaneously, and then the target model verifies all generated tokens in parallel. However, such rigid serialized processing may reduce system throughput under a highly dynamic load, as observed in real-world traces ~\cite{stojkovic2025dynamollm, wang2025burstgpt,patel2024splitwise}.

\begin{figure}[t]
    \centering
    \begin{subfigure}[t]{0.48\linewidth}
        \centering
        \includegraphics[width=\linewidth]{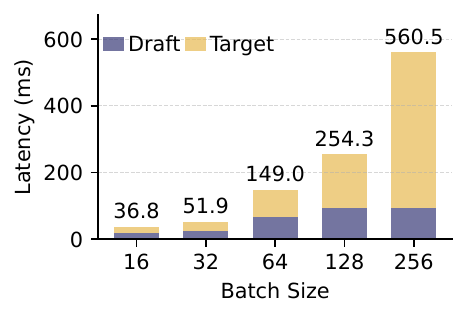}
        \vspace{-1.8em}
        \caption{Absolute latency}
        \label{fig:latency_absolute_breakdown}
    \end{subfigure}
    \hfill
    \begin{subfigure}[t]{0.48\linewidth}
        \centering
        \includegraphics[width=\linewidth]{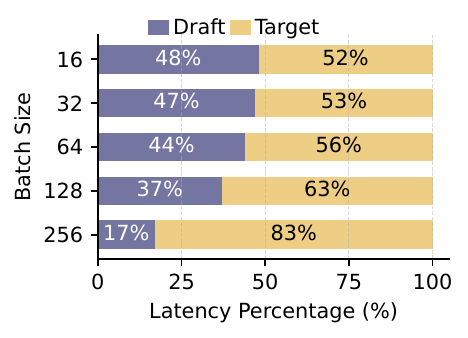}
        \vspace{-1.8em}
        \caption{Latency breakdown}
        \label{fig:latency_breakdown_horizontal}
    \end{subfigure}
    \vspace{-1em}
    \caption{Decode latency breakdown, showing the draft and verification phases' contributions, absolute (a) and relative (b).
    Draft/target model are Qwen3-0.6B/Qwen3-32B.}
    \vspace{-15pt}
    \label{fig:latency_percentages}
\end{figure}

\subsection{SD System Efficiency under Dynamic Workload}
\label{sec:characteristics}
Real-world LLM serving deployments exhibit highly dynamic workloads~\cite{stojkovic2025dynamollm,patel2024splitwise,wang2025burstgpt}, where the valley periods show 1.7$\sim$35$\times$ lower RPS compared to the peak periods~\cite{stojkovic2025dynamollm}. 
Moreover, prior work~\cite{lai2025tokenscale} confirms substantial fluctuations in batch size when replaying these traces in a GPU cluster.
Hence, we study the decode bottlenecks in SD systems across different batch sizes using the \texttt{Chat} trace from DynamoLLM~\cite{stojkovic2025dynamollm} to model request arrivals, ShareGPT~\cite{sharegpt} to provide realistic prompts, Qwen3-0.6B as the draft model, and Qwen3-32B as the target model, all evaluated on a 96GB NVIDIA H100 GPU with a fixed speculative length of 6.

\textbf{Verification becomes the dominant computational bottleneck of SD under high load.}
Fig.~\ref{fig:latency_percentages} shows the inference latency, measured as time per output token, and this latency's breakdown between draft and target latency components in an SD system, when sweeping the batch size from 16 to 256. We observe that overall latency increases significantly with increasing the batch size (Fig.~\ref{fig:latency_absolute_breakdown}). Fig.~\ref{fig:latency_breakdown_horizontal} further shows that verification consistently contributes more latency than drafting, and this dominance becomes more pronounced with larger batch sizes, with verification accounting for up to 83\% of the total latency. The underlying reason is that verification over large batches places much heavier demand on GPU compute resources, saturating them, while draft latency remains comparatively stable because the draft model is substantially smaller and faster.

\textbf{Serial execution of the draft and target phases for small request batches substantially increases the latency.}
With small batch sizes, the draft phase contributes nearly half of the generation latency, as shown in Fig.~\ref{fig:latency_breakdown_horizontal}.
The reason is that most existing speculative decoding systems~\cite{liu2024optimizing,huang2025adaspec,li2025adaserve,miao2024specinfer} execute draft generation and target verification strictly sequentially, blocking the verification phase from starting until draft generation for the entire batch completes. This serialized execution results in poor GPU utilization and longer decode times.

\begin{figure}[t]
    \centering
    \begin{subfigure}[t]{0.48\linewidth}
        \centering
        \includegraphics[width=\linewidth]{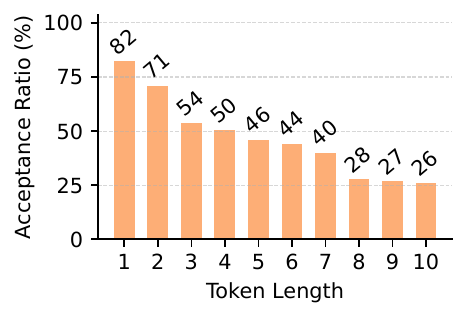}
        \vspace{-1.8em}
        \caption{Acceptance ratio}
        \label{fig:acceptance_ratio_by_token_length}
    \end{subfigure}
    \hfill
    \begin{subfigure}[t]{0.48\linewidth}
        \centering
        \includegraphics[width=\linewidth]{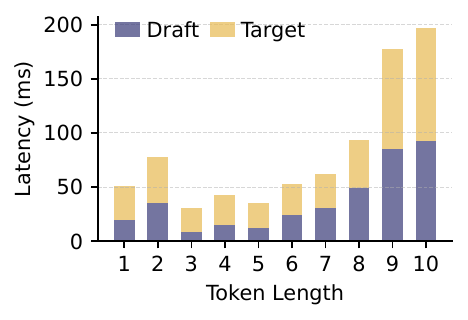}
        \vspace{-1.8em}
        \caption{Latency}
        \label{fig:accept_ratio_by_running_reqs_token_length_6}
    \end{subfigure}
    \vspace{-1em}
    \caption{Acceptance ratio and decode latency when increasing speculative token length, with the batch size of 32.} 
    \vspace{-16pt}
    \label{fig:acceptance_ratio_and_latency_by_token_length}
\end{figure}

\textbf{Higher acceptance ratio does not always lead to lower latency.} Unlike prior works that focus solely on maximizing acceptance ratios~\cite{xia2024unlocking,zhang2024beyond,zhang2024draft,zimmer2024mixture}, we observe that the speculative token length creates a fundamental trade-off between acceptance efficiency and verification overhead.
As shown in Fig.~\ref{fig:acceptance_ratio_by_token_length}, increasing token length naturally decreases the acceptance ratio because later speculative tokens are more likely to deviate and be rejected. However, the highest acceptance ratio (achieved at length 1) does not yield the lowest latency; instead, our setup achieves optimal latency at length 3 (Fig.~\ref{fig:accept_ratio_by_running_reqs_token_length_6}). 
This is because shorter token lengths require more SD iterations, bottlenecking performance with frequent, unamortized verification rounds. Conversely, longer lengths reduce the number of iterations but waste substantial compute: the target model must verify the entire draft sequence in parallel, wasting heavy computational resources on suffix tokens destined for rejection. Ultimately, the cost of verifying these rejected tokens outstrips the benefits of fewer iterations. SD systems must therefore dynamically navigate this trade-off, adjusting token length based on current load and workload characteristics.

\begin{table}[t]
\centering
\small
\setlength{\tabcolsep}{3pt}
\caption{Comparison of representative SD systems in addressing problems when serving dynamic loads.}
\label{tab:motivation_compare}
\vspace{-1em}
\resizebox{\linewidth}{!}{
\renewcommand{\arraystretch}{0.65}
\begin{tabular}{C{2.2cm}C{2.2cm}C{2.6cm}C{1.8cm}}
\toprule
\textbf{Method} &
\begin{tabular}[c]{@{}c@{}}\textbf{Static Spec. }\\\textbf{Length}\end{tabular} &
\begin{tabular}[c]{@{}c@{}}\textbf{High-overhead}\\\textbf{Verification}\end{tabular} &
\begin{tabular}[c]{@{}c@{}}\textbf{Serial}\\\textbf{Execution}\end{tabular} \\
\midrule
SpecInfer~\cite{liu2024optimizing} & \xmark & \cmark & \xmark \\
AdaSpec~\cite{huang2025adaspec} & \cmark &  \xmark & \xmark \\
Smurfs~\cite{wang2025towards} &\cmark & \xmark & \cmark \\
\midrule
\textbf{\sys (Ours)} &\cmark & \cmark & \cmark \\
\bottomrule
\end{tabular}
}
\end{table}

\vspace{-.3em}
\subsection{Limitations of Existing Systems}
\label{sec:limitations}

Recent work has explored various optimizations to alleviate the verification bottleneck and the acceptance-ratio/latency trade-off induced by speculative token length, as discussed in~\cref{sec:characteristics}.
Nevertheless, under dynamic request arrivals in real-world traces, representative SD serving systems still fall short in addressing several key inefficiencies, including the \emph{static speculative length}, \emph{high-overhead verification}, and \emph{serial execution}, as summarized in Table~\ref{tab:motivation_compare}.

\textbf{SpecInfer}~\cite{sun2024spectr} improves SD mainly by increasing candidate diversity and verifying a token tree in a single target pass. While it partially mitigates the constraints of a \emph{static speculative length} via tree-based speculation, this design primarily focuses on reducing the total number of verification rounds rather than the latency of each individual round. Moreover, the computational complexity of tree construction, combined with the strictly \emph{serial execution} of drafting and verification phases, leaves GPU resources underutilized under low load. Under high load, it fails to resolve the \emph{high-overhead verification} bottleneck, as the target model must process the entire tree structure.

\textbf{AdaSpec}~\cite{huang2025adaspec} addresses the \emph{static speculative length} through adaptive tuning and confidence-guided token elimination with considering SLO constraints. However, it fails to optimize the \emph{high-overhead verification} stage because the target model still verifies the entire speculative sequence. Consequently, it cannot avoid wasting substantial computation on tokens in the rejected suffix, which remains a critical path bottleneck. Furthermore, AdaSpec maintains a \emph{serial execution} workflow, which results in poor GPU utilization and prolonged latency under dynamic, low-load traffic.

\textbf{Smurfs}~\cite{wang2025towards} optimizes mixed-task scenarios by leveraging multiple SSMs and adapts speculation length online to balance the accepted tokens against verification cost, thereby addressing the \emph{static speculative length} issue. It further introduces pipelined execution to overlap SSM speculation with LLM verification across batches which addressed \emph{serial execution}. However, this overlap primarily operates at a coarse granularity and fails to exploit fine-grained spatial multiplexing on a single GPU. Consequently, Smurfs falls short in fundamentally reducing the \emph{high-overhead verification} work, as the selected speculative tokens are still verified in a round-based manner without the ability to prune the verification critical path mid-execution.

The limitations of these existing approaches, particularly their inability to adapt verification behavior and serial execution of drafting and verification to dynamic loads, motivate a more holistic SD system \sys. By addressing all three key inefficiencies in a unified design, \sys responsively adapts to dynamic workloads through the considering both draft-target interaction and verification behavior.

\vspace{-.3em}
\section{Opportunities and Challenges}
\label{sec:opportunities_and_challenges}
Based on the above observations, we identify two opportunities and the corresponding challenges for re-designing SD systems to serve highly dynamic LLM inference traffic.

\begin{figure}[t]
  \vspace{-1em}
  \centering
  \includegraphics[width=\linewidth]{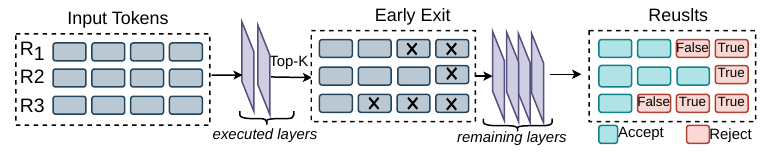}
  \vspace{-2em}
  \caption{Example of the token-wise early-exit method. A token marked with $\times$ is exited early and does not participate in the remaining layers. \texttt{True} and \texttt{False} in the box indicate whether the early-exit decision agrees with the outcome of full verification without early exit.}
  \vspace{-10pt}
  \label{fig:early-exit}
\end{figure}

\textbf{\textit{Opportunity 1}: Early Exit to Combine Deep Speculation with Low Verification Overhead.}
\label{sec:opportunity1}
In~\cref{sec:characteristics}, we show that increasing the acceptance ratio may not always improve performance due to the increasing computational overhead of verifying the tokens from the rejected suffix. For example, with a batch size of 32 and a speculative length of 6, we see, on average, more than half of the tokens are rejected during verification, meaning that around half of the FLOPS spent on inference serving are wasted.
This observation creates an opportunity: if the system could predict the boundary between accepted tokens and tokens in the rejected suffix, it could terminate early, skipping the computation for the rejected suffix.

To exploit this opportunity, we explore a token-wise early-exit mechanism inspired by prior early-exit methods for standard autoregressive decoding~\cite{fan2024not}. 
As illustrated in Fig.~\ref{fig:early-exit}, after each verification layer, the system uses the intermediate outputs to estimate whether further verification of the earliest unverified drafted token is worthwhile. If not, we terminate verification early and avoid executing deeper layers for that token and its speculative suffix. Otherwise, we continue verification to the next layer and repeat the same decision process. In this way, the verifier can adaptively skip unnecessary work for low-value speculative continuations while still fully processing promising ones.

\begin{figure}[t]
    \centering
    \begin{subfigure}[t]{0.48\linewidth}
        \centering
        \includegraphics[width=\linewidth]{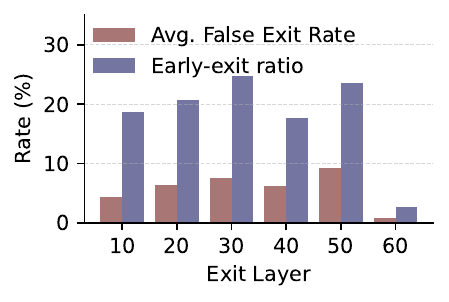}
        \vspace{-1.8em}
        \caption{Predictor accuracy}
        \label{fig:early_exit_ratio_analysis}
    \end{subfigure}
    \hfill
    \begin{subfigure}[t]{0.48\linewidth}
        \centering
        \includegraphics[width=\linewidth]{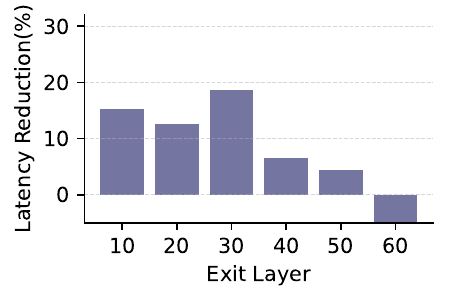}
        \vspace{-1.8em}
        \caption{Inference latency reduction}
        \label{fig:early_exit_benefit_analysis}
    \end{subfigure}
    \vspace{-1.2em}
    \caption{Accuracy of the rejected suffix predictor, shown as total and false early exit rates at different layers, and the associated latency reduction, measured with the speculative length of 6 and batch size of 32.}
    \label{fig:acceptance_ratios}
\end{figure}

The key \textbf{challenge ($C_1$)} to enable efficient early exit is to accurately predict reject suffix with low overhead.
Using the same experimental setup as in \cref{sec:characteristics}, we find that a simple Top-$K$ ($K{=}10$) logit-based signal between the draft and target models can predict the beginning of the rejected suffix with high accuracy, reaching up to 95\%, at several intermediate layers of the target model, before full verification completes.
This signal allows the verifier to bypass a substantial fraction 25\% of verification computation and yields up to 19\% latency reduction, as shown in Fig.~\ref{fig:early_exit_ratio_analysis} and Fig.~\ref{fig:early_exit_benefit_analysis}. 
However, SD systems do not directly expose the interaction between the draft and target models, requiring additional instrumentation in the runtime. 
Moreover, the choice of $K$ directly affects the tradeoff between prediction accuracy and signal-computation overhead. As a result, designing an effective early-exit mechanism requires jointly considering the prediction method and its control parameters, which becomes even more difficult with dynamic workloads~(\cref{sec:characteristics}).

\begin{figure}[t]
  \vspace{-0.5em}
  \centering
  \includegraphics[width=\linewidth]{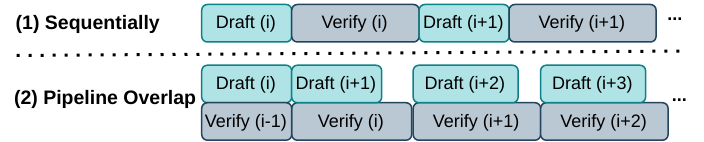}
  \vspace{-1.5em}
  \caption{Example of pipeline overlap between draft generation and target verification. $i$ means the $i$-th iteration of draft generation or target verification in SD.}
  \label{fig:pipeline_draft_target}
  \vspace{-15pt}
\end{figure}

\textbf{\textit{Opportunity 2:} Latency Reduction through Fine-grained Draft-Verification Overlap.}
\label{sec:opportunity2}
In existing SD systems, each decoding iteration is still executed largely as two serialized stages on the same GPU: draft generation followed by target verification. This stage-by-stage execution can unnecessarily increase latency, because neither stage fully utilizes the GPU in isolation. In our experiments with a batch size of 128 and a speculative length of 6, we still observe SM occupancy below 20\%, indicating that substantial GPU resources remain idle even under a relatively large batch. Moreover, the draft stage typically exhibits much lower SM occupancy than target-side verification~\cite{lu2026dfvg}. These results suggest that the two stages need not execute in strict isolation, and that the unused hardware slack can be exploited to overlap them.

This underutilization creates an opportunity for fine-grained pipelining across SD iterations to reduce the latency. As illustrated in Fig.~\ref{fig:pipeline_draft_target}, the draft execution of iteration $i$ can overlap with the target verification of iteration $i\!-\!1$, allowing the system to exploit otherwise idle GPU resources and reduce end-to-end latency.

However, realizing this opportunity is non-trivial because draft generation and target verification are strongly coupled in SD. The draft model cannot continue generating arbitrarily far ahead: whether subsequent drafting is valid depends on the verification result of previously drafted tokens. Once verification rejects a token, all later speculative tokens become invalid and must be discarded, so the next drafting step must restart from the verified prefix. As a result, draft generation is gated by verification progress, making overlap fundamentally different from simply co-scheduling two independent GPU tasks. This leads to another key \textbf{challenge}~($C_2$): enabling fine-grained overlap without violating the strict dependency between verification outcomes and subsequent drafting. Under dynamic workloads, changing batch composition and acceptance behavior can continuously shift the valid drafting frontier, making static overlap policies ineffective.

%% file: sections/3-design.tex
\section{System Design}
\label{sec:design}

\begin{figure}[t]
  \centering
  \includegraphics[width=\linewidth]{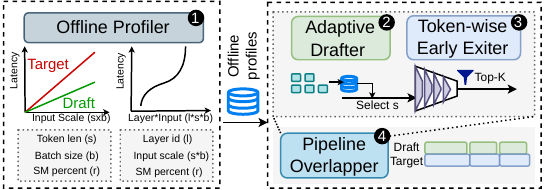}
  \vspace{-1em}
  \caption{\sys architecture overview.}
  \label{fig:system_overview}
  \vspace{-15pt}
\end{figure}

Motivated by the opportunities and challenges identified in~\cref{sec:opportunities_and_challenges}, we develop \sys, a system that jointly optimizes the draft and verify stages to reduce per-token inference latency and improve overall throughput under dynamic loads.

Fig.~\ref{fig:system_overview} illustrates the overall workflow and key design of \sys. At a high level, \sys serves incoming inference requests through a controller that jointly determines (1) how many speculative tokens to draft, (2) how far each drafted token should proceed in verification, and (3) how GPU SM resources should be partitioned between the draft and target models to improve pipeline efficiency.

To support these online decisions, the \textit{\textbf{Offline Profiler}} \ding{182} first profiles the draft-target pair under a range of execution conditions, including different batch sizes, speculative lengths, and SM partition configurations. It records the latency and throughput characteristics of both models, as well as the performance impact of overlapping their execution under different resource splits. These profiling results are then summarized into lightweight performance tables used by the online controller to predict the latency and the benefit of different serving decisions.

Inference requests arrive continuously and are first handled by the \textit{\textbf{Adaptive Drafter~\ding{183}}}. Based on the current batch size, request load, and offline profiling results, it determines the speculative length for the current verification round. The objective is to select a draft length that provides enough speculative parallelism to improve efficiency, while avoiding overly long drafts that would increase verification overhead and reduce acceptance efficiency.

The drafted tokens are then processed by the \textit{\textbf{Token-wise Early Exiter~\ding{184}}} during target-side verification. At each verification layer, \sys evaluates whether each drafted token remains worthwhile to continue verifying based on its Top-$K$ signal and the current system condition. Once a token is judged unlikely to be accepted, \sys terminates verification for that token and discards its speculative suffix, thereby avoiding unnecessary deeper-layer computation and reducing verification time.

To further maximize efficiency, the \textbf{\textit{Pipeline Overlapper~\ding{185}}} enables fine-grained overlap between draft generation and target verification, transforming these traditionally serialized stages into a concurrent pipeline. By leveraging GreenContexts~\cite{greencontext}, \sys explicitly partitions GPU SM resources to co-locate the draft-target pair with minimal resource interference. This hardware-aware spatial multiplexing effectively reduces pipeline bubbles and exploits idle hardware slack, leading to significantly improved GPU utilization and overall serving throughput.

\vspace{-.3em}
\subsection{Adaptive Drafting}
\label{sec:adaptive_drafting}
Being fully compatible with continuous batching~\cite{kwon2023efficient}, \sys chooses speculative length for newly arrived requests at run time while adjusting it for those that are already in a batch. This fine-grained adaptivity allows \sys to tailor to the dynamic workload characteristics in contrast to the prior methods~\cite{huang2025specserve,li2025nightjar,wu2025tetris,liu2024speculative,svirschevski2024specexec}, which choose speculative length for all requests in an entire batch despite the growing diversity in decode requests, e.g., coding, conversation, agentic, reasoning, and highly volatile inter-arrival distribution, like in BurstGPT~\cite{wang2025burstgpt}. 

The \textit{\textbf{Adaptive Drafter}} in \sys jointly considers acceptance behavior and runtime latency under the current batch size $b$ and SM allocation $r$, and assigns each request $i$ its own speculative length $s_i$.
However, choosing the speculative length for each request is still non-trivial. A larger speculative length may increase the number of committed tokens when the acceptance ratio is high, but it also increases verification latency. Moreover, although the speculative length is selected per request, all requests are still executed under the same batch context and therefore remain coupled through the current batch size $b$ and SM allocation $r$. As a result, the best speculative length for each request should be chosen jointly according to its acceptance behavior and runtime latency under the current $(b,r)$.

\subsubsection{Optimization Objective}
For each request $i$ and each candidate speculative length $s \in \mathcal{S}$, the system maintains two online statistics from recent batches under the current serving context $(b,r)$: the historical acceptance estimate $\hat{a}_i(s;b,r)$ and the observed batch-latency estimate $\hat{T}(s;b,r)$. We use them to estimate the average latency contribution per committed token:
\vspace{-.7em}
\begin{equation}
  \small
  J_i(s;b,r)=
  \frac{\hat{T}(s;b,r)}
  {s \cdot \hat{a}_i(s;b,r)+\epsilon},
  \label{eq:length_objective}
  \vspace{-.3em}
\end{equation}
where $\mathcal{S}$ denotes the discrete candidate set of speculative lengths and $\epsilon$ is a small constant for numerical stability. Here, $s \cdot \hat{a}_i(s;b,r)$ is the expected number of committed tokens contributed by request $i$ in one speculative verification step under the current context $(b,r)$. The controller then assigns request $i$ the speculative length
\vspace{-.5em}
\begin{equation}
  \small
  s_i^*(b,r)=\arg\min_{s\in\mathcal{C}} J_i(s;b,r).
  \label{eq:opt_length}
  \vspace{-.3em}
\end{equation}
Although latency is observed at batch granularity and therefore shared by requests in the same verification step, we treat $\hat{T}(s;b,r)$ as a contextual feedback signal under the current $(b,r)$, while keeping the acceptance term request-specific through $\hat{a}_i(s;b,r)$.

This objective is simple but effective. A higher acceptance ratio reduces the cost by increasing useful committed tokens, while a higher latency increases the cost. As a result, the selected $s$ naturally balances the benefit of longer speculation against its additional verification overhead.

\subsubsection{GP-LCB-Based Online Search}
\label{sec:gplcb}
Although $\mathcal{S}$ is small, probing all speculative lengths online is still undesirable because each trial affects serving latency and the objective in Eq.~\ref{eq:length_objective} is noisy under dynamic loads.
We therefore treat $J(s;b,r)$ as a black-box function and use GP-LCB~\cite{cox1992statistical} to search for its minimizer.

At round $n$, the \textit{\textbf{Adaptive Drafter}} maintains a Gaussian Process posterior with mean $\mu_{n-1}(s;b,r)$ and standard deviation $\sigma_{n-1}(s;b,r)$ for each candidate $s \in \mathcal{S}$.
The next speculative length is selected as
\vspace{-.7em}
\begin{equation}
s_n=
\arg\min_{s\in\mathcal{S}}
\mu_{n-1}(s;b,r)-\sqrt{\beta_n}\,\sigma_{n-1}(s;b,r),
\label{eq:gplcb}
\vspace{-.5em}
\end{equation}
where $\beta_n$ balances exploitation and exploration.
After executing the batch with $s_n$, the \textit{\textbf{Adaptive Drafter}} records the observed cost
\vspace{-.9em}
\begin{equation}
\small
\tilde{J}(s_n;b,r)=
\frac{T^{\mathrm{obs}}(s_n,b,r)}
{s_n \cdot a^{\mathrm{obs}}(s_n,b,r)+\epsilon},
\label{eq:realized_cost}
\vspace{-.5em}
\end{equation}
and uses it to update the Gaussian Process posterior.

To remain adaptive, the \textit{\textbf{Adaptive Drafter}}  maintains statistics over a sliding window of recent batches rather than the full history. This design allows it to quickly respond to changes in load, batch size, and available SM resources, while avoiding repeated sweeps over the full candidate set.

\begin{figure}
    \centering
    \includegraphics[width=\linewidth]{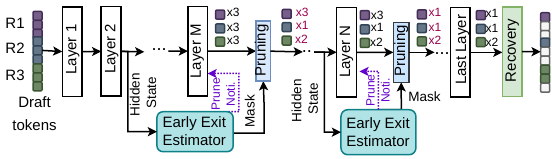}
    \vspace{-1.2em}
    \caption{The workflow of adaptive token-wise early exiting. Input includes three requests, each with 3 draft tokens. The output gray blocks represent the token with no logits.}
    \label{fig:token_wise_early_exit_verfi}
    \vspace{-1.5em}
\end{figure}

\vspace{-.5em}
\subsection{Token-wise Early Exiting}
\label{sec:token-wise-early-exit}
When drafted tokens are submitted to the target model for verification, \sys employs a lightweight token-wise early-exit mechanism to avoid unnecessary verification on tokens that would ultimately be rejected, thus reducing the wasted verification work highlighted by challenge $C_1$ (\cref{sec:opportunity1}).

As shown in Fig.~\ref{fig:token_wise_early_exit_verfi}, \sys enables token-wise early exit after the first $L_{\mathrm{init}}$ layers, where hidden states become sufficiently informative for verification pruning decisions. Starting from layer $\ell \ge L_{\mathrm{init}}$, the hidden states of the currently active speculative tokens are forwarded to the estimator asynchronously, while the main verification stream continues without blocking. When the estimator returns an updated mask, the next transformer layer immediately prunes the tokens predicted to belong to the rejected suffix, and only the remaining tokens proceed to subsequent layers. This process repeats iteratively across layers until verification completes. Finally, before logits are produced, \sys restores pruned positions to the original tensor layout to preserve compatibility with the downstream speculative-decoding pipeline.

To improve pruning efficiency while minimizing the re-drafting iteration caused by the incorrect pruning, \sys adopts an acceptance-aware pruning policy coordinated with the \textit{\textbf{Adaptive Drafter}}. Recall that each request $i$ in the current batch can use its own speculative length $s_i$ under the same serving context $(b,r)$. For request $i$, the estimator uses the historical acceptance estimate $\hat{a}(s_i;b,r)$ to approximate the number of speculative tokens that are likely to fall into the rejected suffix:
\vspace{-.5em}
\begin{equation}
  \small
  \hat{k}_{\mathrm{rej},i}
  =
  s_i \cdot \bigl(1-\hat{a}(s_i;b,r)\bigr).
  \label{eq:rej_token_req}
  \vspace{-.5em}
\end{equation}
Aggregating over the current batch $\mathcal{B}$, the estimated number of potentially prunable tokens is
\vspace{-.5em}
\begin{equation}
  \small
  \hat{k}_{\mathrm{rej}}(\mathcal{B};b,r)
  =
  \sum_{i\in\mathcal{B}} \hat{k}_{\mathrm{rej},i}.
  \label{eq:rej_token_batch}
  \vspace{-.5em}
\end{equation}

Let $T_t(b,r,s)$ denote the measured target-side verification time for $s$ speculative tokens under batch size $b$ and SM allocation $r$, $T_{ee}(b,r,s)$ the latency of the early-exit estimation, and $T_{pr}(b,r,s)$ the pruning overhead.
If the estimator is invoked after layer $\ell$, the remaining verification depth is $ L_r(\ell)=L-\ell.$
Because the estimator runs asynchronously, its latency does not directly stall the main stream, but it reduces the remaining layers over which pruning can still save work. We therefore convert estimator latency into an equivalent number of consumed verification layers:
\vspace{-.5em}
\begin{equation}
  \small
  \Delta L_{ee}(b,r,s)
  =
  \frac{L \cdot T_{ee}(b,r,s)}{T_t(b,r,s)}.
  \label{eq:lee}
  \vspace{-.5em}
\end{equation}
The effective remaining depth that can still benefit from pruning is then
\vspace{-.5em}
\begin{equation}
  \small
  L_r'(\ell;b,r,s)
  =
  \max\!\bigl(L_r(\ell)-\Delta L_{ee}(b,r,s),\,0\bigr).
  \label{eq:remaining_layer_effective}
  \vspace{-.5em}
\end{equation}
which gives the maximum verification time that pruning can still save:
\vspace{-.5em}
\begin{equation}
  \small
  T_{\mathrm{save}}(\ell;b,r,s)
  =
  \frac{L_r'(\ell;b,r,s)}{L}\cdot T_t(b,r,s).
  \label{eq:save_time}
  \vspace{-.5em}
\end{equation}

\sys triggers pruning only when the remaining benefit exceeds the pruning overhead under the current serving condition:
\vspace{-.5em}
\begin{equation}
  \small
  T_{\mathrm{save}}\!\bigl(\ell;b,r,\hat{k}_{\mathrm{rej}}(\mathcal{B};b,r)\bigr)
  >
  T_{pr}\!\bigl(b,r,\hat{k}_{\mathrm{rej}}(\mathcal{B};b,r)\bigr).
  \label{eq:prune_trigger}
  \vspace{-.5em}
\end{equation}
Otherwise, the estimator output is ignored because the residual verification work is too small to amortize the computational cost of pruning. 

For the token-level decision, \sys follows the Top-$K$ criterion in~\cite{fan2024not}: a drafted token that falls outside the current Top-$K$ set of the intermediate logits is marked as unlikely to be accepted and becomes a pruning candidate. To reduce erroneous pruning in shallow layers, \sys uses a depth-dependent threshold $K_\ell$, which is conservative near $L_{\mathrm{init}}$ and becomes stricter only at deeper layers when hidden states are more reliable. Regardless of whether pruning is triggered, the current layer output is still forwarded to the estimator for the next asynchronous early-exit evaluation.

\begin{figure}[t]
    \centering
    \includegraphics[width=\linewidth]{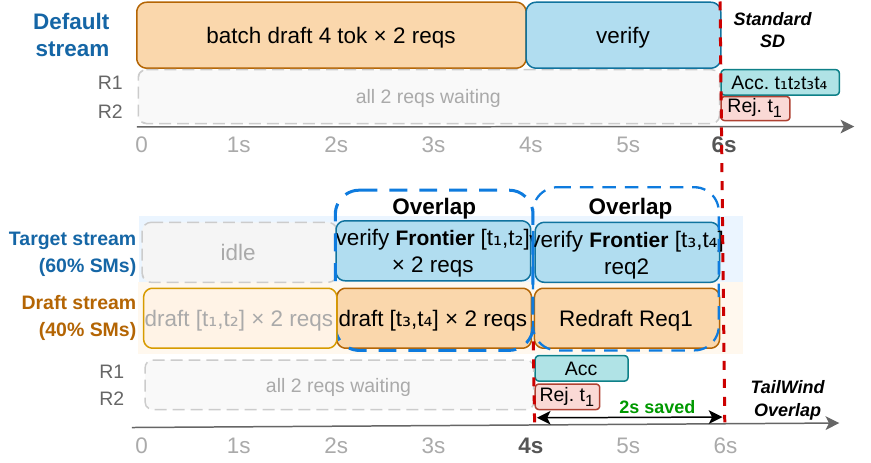}
    \vspace{-1.5em}
    \caption{Illustration of overlapping draft and verification phases in \sys. The speculative length is set as 4, and the frontier chunk size is 2.
    } 
    \vspace{-1.8em}
    \label{fig:pipeline_overlap}
\end{figure}

\vspace{-.3em}
\subsection{Fine-grained Overlap of the SD Phases}
\label{sec:pipeline_overlapping}
\sys aims to transform draft-verification overlap into a practical mechanism for SD, with addressing $C_2$ highlighted in \cref{sec:opportunity2}. The design must satisfy three requirements: expose overlap at fine granularity rather than as two monolithic stages, ensure that the latency hidden by co-execution outweighs the interference it introduces, while preserving the original correctness semantics of speculative decoding.

To this end, \sys organizes speculative tokens as a \emph{frontier}, i.e., the currently verifiable prefix of drafted tokens, and lets the target model verify this frontier incrementally in chunks. Once a frontier chunk is verified, accepted tokens are committed immediately. If a rejection is observed, \sys resets the frontier: it commits the accepted prefix together with the recovery token produced by the target model, discards all remaining speculative tokens beyond the rejection point, and restarts drafting from the newly committed prefix. This reset-and-continue behavior allows \sys to sustain speculative work within the same scheduling step, so that a rejection does not introduce an idle gap before useful drafting resumes. In this way, frontier execution changes when speculative work is exposed and consumed, but not the correctness semantics of speculative decoding.

Based on this frontier abstraction, \sys overlaps drafting and verification at chunk granularity. As shown in Fig.~\ref{fig:pipeline_overlap}, \sys places the draft and target models in separate Green Contexts with controlled SM allocations, allowing them to make progress concurrently with bounded compute-side interference. While the target model verifies the current frontier chunk, the draft model generates the next chunk in parallel using the latest committed prefix, so the system no longer alternates between a full drafting phase and a full verification phase. At each synchronization point, newly drafted tokens are merged into the active verification buffer, and the next verification round begins immediately. This turns speculative decoding into a finer-grained producer-consumer pipeline, in which verification consumes the current frontier chunk while drafting prepares the next one in parallel. The chunk size $s_c$ is chosen to balance draft and verification stages according to their estimated execution times, according to the requirement: $
    T_q(s_c;b,r)+\frac{s}{s_c} \cdot T_p(s_c;b,1-r) < \hat{T}(s;b,r),$
where $T_q$ and $T_p$ are the draft and target latency with given batch size $b$, SM allocations $r$($1-r$), and speculative length $s$.

When a chunk is fully accepted, the pre-drafted next chunk is available, and verification can continue without delay, as illustrated by R2 in Fig.~\ref{fig:pipeline_overlap}. When a chunk is rejected, the frontier is reset: all speculative tokens beyond the rejection point are discarded, including any generated for subsequent chunks, and drafting restarts from the newly committed prefix. Thus, overlap changes only the execution schedule, not the acceptance behavior of speculative decoding.

Together, frontier execution and Green Contexts make overlap practical. Frontier execution defines when speculative work becomes verifiable, when to commit it, and when to discard it after rejection. Green Contexts bound compute-side interference while allowing both models to progress concurrently, though shared HBM bandwidth and cache hierarchy still introduce memory contention. \sys therefore enables overlap selectively, only when the draft latency savings are expected to outweigh co-execution overhead.

\vspace{-.3em}
\subsection{Offline Profiling}
\label{sec:offline_profiling}
To support \sys's mechanisms, \sys performs offline profiling of the draft model, target model, and token-wise early-exit verification latency under different batch sizes, token lengths, and SM allocations. The resulting profiles guide dynamic parameter adjustment during online serving. \sys runs all profilers in a daemon that updates their parameters in the background.

\subsubsection{Draft Latency Profiling.}
\label{sec:draft_profiling}
The latency of the draft model is mainly affected by three factors: batch size $b$, token length $s$, and SM allocation $r$. 
We observe that, for fixed $b$ and $s$, the draft latency varies approximately piecewise linearly with the allocated SMs $r$. Intuitively, increasing $r$ reduces draft latency, but the marginal benefit is not uniform across the full allocation range due to changes in kernel occupancy and resource contention. We therefore approximate the draft latency $T_q$ using the piecewise model:
\vspace{-.9em}
\begin{equation} 
  \small
  T_q(\star)=\begin{cases} 
    (a_{q,1}-\gamma_{q,1} r)\cdot(\alpha_q b+\beta_q s+c_q), & 0 < r \le R_q,\\[4pt] 
    (a_{q,2}-\gamma_{q,2} r)\cdot(\alpha_q b+\beta_q s+c_q), & R_q < r \le 1. 
  \end{cases} 
  \label{eq:draft_latency_model}
  \vspace{-.5em} 
\end{equation}
Here, $\star$ denotes the input tuple $(b,s,r)$, $R_q$ is the SM allocation threshold at which the slope changes, and $a_q$, $\gamma_{q,1}$, $\gamma_{q,2}$, $\alpha_q$, $\beta_q$, and $c_q$ are fitted based on the profiling data. This model captures the piecewise-linear dependence of latency on SM allocation while maintaining a lightweight form for efficient online estimation.

We use this model because prior work~\cite {chen2025multiplexing,choi2022serving} has shown that latency often exhibits a piecewise-linear trend with respect to SM allocation. In particular, the latency curve typically has a knee point where the slope changes, reflecting the underlying GPU architecture and the model's resource demands. The model parameters can be efficiently fitted using standard regression techniques over the offline profiling data. We discuss the fitting accuracy in~\ref{sec:accuracy_offline_profiling}.

\subsubsection{Target Latency Profiling.}
\label{sec:target_profiling}
Unlike the draft model, the target model has a much larger parameter size, making its inference latency primarily dominated by computational load and memory access overhead. In Transformer-based models, the dominant operations are GEMMs~\cite{lee2025forecasting}, whose total floating-point operations scale with the product of batch size $b$ and token length $s$, i.e., $O(b \cdot s \cdot d)$, where $d$ is the model dimension. Consequently, the target latency exhibits a near-linear dependence on $b \cdot s$.

However, the increase in latency is not strictly linear with respect to the speculative length.
As batch size increases, the target model typically achieves better GPU utilization, which changes how latency grows with $s$. Moreover, under spatial partitioning, if the draft model is allocated with $r$ fraction of SMs,then the target model receives the remaining share of $1-r$, which also affects its latency. Similar to draft latency profiling, we observe that, for fixed $b$ and $s$, the target inference latency approximately follows a piecewise linear function with a $1-r$ fraction of SMs. We therefore model the target latency $T_p$ as
\vspace{-.5em}
\begin{equation}
  \small
T_p(\star)\!=\!
\begin{cases}
(a_{p,1}\!-\!\gamma_{p,1}(1\!-\!r))\,((\alpha_p b\!+\!\lambda_p)s\!+\!c_p), & r \ge 1\!-\!R_p,\\[2pt]
(a_{p,2}\!-\!\gamma_{p,2}(1\!-\!r))\,((\alpha_p b\!+\!\lambda_p)s\!+\!c_p), & r < 1\!-\!R_p.
\end{cases}
\label{eq:target_model_exec_time}
\vspace{-.3em}
\end{equation}
where $r$ is the SMs allocated to the draft model, and $a_p$, $\gamma_{p,1}$, $\gamma_{p,2}$, $\alpha_p$, $\lambda_p$, and $c_p$ are fitted from the profiling data. This model captures the piecewise-linear dependence of target latency on SM allocation while also accounting for the linear scaling with batch size and token length.

\subsubsection{Early-Exit Latency Profiling.}
\label{sec:early_exit_profiling}
As discussed in~\cref{sec:token-wise-early-exit}, the early-exit overhead consists of two parts: the latency of early-exit estimation and the latency of pruning the estimated rejected suffix' tokens. The checking stage converts hidden states into logits and determines the Top-$K$ candidates, while the pruning stage removes tokens rejected by early exit.

Under a fixed model configuration, the latency of both stages grows approximately linearly with the number of processed tokens. Under spatial partitioning, however, the two stages are constrained by different SM allocations: early-exit estimation runs on the target side and is therefore governed by the target-side share, \(1-r\), whereas pruning operates on drafted tokens and is therefore governed by the draft-side share, \(r\). Similar to the draft and target latency models, we represent both costs using a linear term in \(b \cdot s\) together with a piecewise-linear factor that captures the effect of SM allocation.
Therefore, we model the early-exit checking latency as
\vspace{-.3em}
\begin{equation}
\small
T_{ee}(\star)\!=\!
\begin{cases}
(a_{ee,1}\!-\!\gamma_{ee,1}(1\!-\!r))\cdot(\alpha_{ee}bs\!+\!\beta_{ee}), & r \ge 1\!-\!R_{ee},\\[4pt]
(a_{ee,2}\!-\!\gamma_{ee,2}(1\!-\!r))\cdot(\alpha_{ee}bs\!+\!\beta_{ee}), & r < 1-R_{ee},
\end{cases}
\label{eq:early_exit_check_time}
\end{equation}
and the pruning latency as
\begin{equation}
\small
T_{pr}(\star)=
\begin{cases}
(a_{pr,1}-\gamma_{pr,1}r)\cdot(\alpha_{pr}bs+\beta_{pr}), & 0 < r \le R_{pr},\\[4pt]
(a_{pr,2}-\gamma_{pr,2}r)\cdot(\alpha_{pr}bs+\beta_{pr}), & R_{pr} < r \le 1.
\end{cases}
\label{eq:pruning_time}
\vspace{-.3em}
\end{equation}
Here, $R_{ee}$ and $R_{pr}$ denote SM-allocation thresholds at which slope changes, and all coefficients are fitted from offline profiling data. These two models keep online estimation lightweight while capturing the dependence of early-exit overhead on both token volume and spatial resource allocation.

%% file: sections/4-implementation.tex
\section{Implementation}
\label{sec:implementation}

We implement \sys on top of vLLM~\cite{vllm} v0.15.1 with 5k lines of Python. We leverage \texttt{PyTorch}, \texttt{numpy}, \texttt{scipy.optimize}, and CUDA Green Contexts via \texttt{cuda-python} across our core components:

\noindent \textbf{Profiling \& Adaptive Drafting:} \sys profiles draft--target pairs across SM allocations, batch sizes, and speculative lengths to build latency and acceptance models, which are refreshed every two hours in a separate process. Online, the runtime controller dynamically selects the speculative length $k$ and drafter SM allocation using a GP-LCB-based search to maximize predicted end-to-end gain.

\noindent \textbf{Token-wise Early Exiting:} The early-exit estimator runs asynchronously on a dedicated \texttt{torch.cuda.Stream}. To minimize memory-bandwidth overhead, \sys avoids full Top-$K$ materialization; it merely checks if at least $k$ tokens exceed the drafted token's probability. Pruned tensors are subsequently padded back to their original shapes to maintain downstream sampling compatibility.

\noindent \textbf{Pipeline Overlapping:} Drafter and verifier executions overlap using separate CUDA streams. To prevent overhead from continuous context creation and destruction, \sys pre-allocates a Green Context pool for SM partitioning and selects configurations dynamically.

%% file: sections/5-methodology.tex

    

%% file: sections/6-evaluation.tex
\section{Evaluation}
\label{sec:evaluation}
We evaluate the performance of \sys across different models and datasets, to demonstrate its effectiveness in reducing the latency and improving the throughput.

\vspace{-.5em}
\subsection{Methodology}
\label{sec:methodology}

\begin{table}[t]
  \centering
  \caption{Model pairs and hardware used in the experiments.}
  \renewcommand{\arraystretch}{0.7}
  \label{tab:model_pairs}
  \vspace{-1em}
  \resizebox{\linewidth}{!}{
  \begin{tabular}{cccc}
      \toprule
      \textbf{Draft Model} & \textbf{Target Model} & \textbf{Hardware(VRAM)} & \textbf{TP} \\
      \midrule
      Qwen3-0.6B & Qwen3-32B & 1$\times$H100 (96GB) & TP=1 \\
      Llama3.2-1B  & Llama3.3-70B & 2$\times$H100 (192GB) & TP=2\\
      \bottomrule
  \end{tabular}}
  \vspace{-0.5em}
\end{table}

\noindent\textbf{Testbed.}
Our experiments run on a server with two NVIDIA H100 GPUs, each with 96\,GB of memory, interconnected via PCIe 4.0. The server uses CUDA 13.1 and NVIDIA Driver 590.48.01, and is equipped with a 64-core AMD EPYC Milan CPU and 256\,GB of host memory.

\noindent\textbf{Models and datasets.}
We evaluate \sys on two model families, Qwen3~\cite{yang2025qwen3} and Llama3~\cite{grattafiori2024llama}, as summarized in Table~\ref{tab:model_pairs}. For the Llama3-70B pair, we deploy the model with tensor parallelism across two GPUs.
Following DistServe~\cite{zhong2024distserve}, we use three datasets: ShareGPT~\cite{sharegpt}, LongBench~\cite{bai2023longbench}, and HumanEval~\cite{li2022evaluating}. The average input/output lengths are 755/200 for ShareGPT, 1738/90 for LongBench, and 171/98 for HumanEval.

\noindent\textbf{Baselines.} 
We compare \sys against three representative speculative decoding baselines: 
\textbf{SpecInfer}~\cite{miao2024specinfer}, \textbf{AdaSpec}~\cite{huang2025adaspec}, and \textbf{Smurfs}~\cite{wang2025towards}; see \cref{sec:limitations} for details.

\noindent\textbf{Request generation.}
We generate dynamic request streams and the request arrival pattern is derived from the Azure LLM invocation trace used by DynamoLLM~\cite{stojkovic2025dynamollm,azuredataset}. We run each configuration five times for reliability. Each run lasts 60 seconds, with an average arrival rate of 26 req/s.

\vspace{-.5em}
\subsection{End-to-End Performance}
\label{sec:e2e_performance}

\begin{figure}[t]
    \centering
    \begin{subfigure}[t]{0.48\linewidth}
        \centering
        \includegraphics[width=\linewidth]{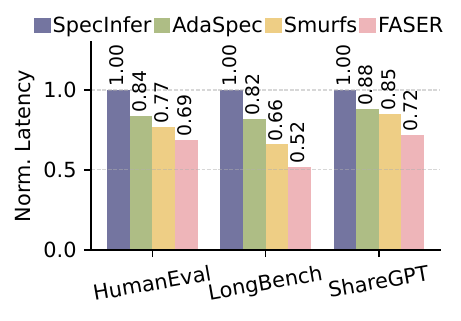}
        \vspace{-1.8em}
        \caption{Qwen3}
        \label{fig:e2e_latency_Qwen3}
    \end{subfigure}
    \hfill
    \begin{subfigure}[t]{0.48\linewidth}
        \centering
        \includegraphics[width=\linewidth]{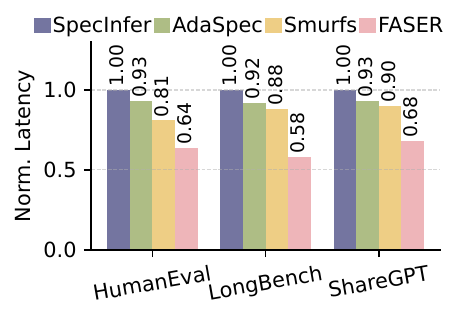}
        \vspace{-1.8em}
        \caption{Llama3}
        \label{fig:e2e_latency_Llama3}
    \end{subfigure}
    \vspace{-1.2em}
    \caption{Latency performance of \sys.}
    \vspace{-1.5em}
    \label{fig:e2e_latency}
\end{figure}

\subsubsection{Latency.}
\label{sec:latency}
We first evaluate the latency of \sys against all baselines. As shown in Fig.~\ref{fig:e2e_latency}, \sys consistently achieves the lowest latency across all models and datasets. For Qwen3, \sys reduces latency by up to 48\% on LongBench. For Llama3, \sys achieves up to 42\% latency reduction on LongBench.
The latency improvement comes from shortening the per-token critical path. Specifically, \sys combines early exit with explicit overlap between draft generation and target verification, so each accepted token requires less target-side work while waiting less for the next verification result. This benefit is particularly pronounced on long-context workloads, where verification overhead is higher and long-tail tokens more easily accumulate on the critical path.
The baselines are less effective because they leave key sources of per-token delay largely intact. SpecInfer incurs additional tree construction and tree verification overhead, while draft generation and target verification remain serialized. AdaSpec designs dynamic length tuning, but still follows a round-based speculate-then-verify workflow and therefore cannot hide draft latency behind target execution. Smurfs introduces pipeline overlap, but it does not directly shorten the target-side verification path that dominates token latency. In contrast, \sys reduces both verification work and cross-stage waiting on the critical path, leading to consistently lower latency across configurations.

\begin{figure}[t]
    \centering
    \begin{subfigure}[t]{0.48\linewidth}
        \centering
        \includegraphics[width=\linewidth]{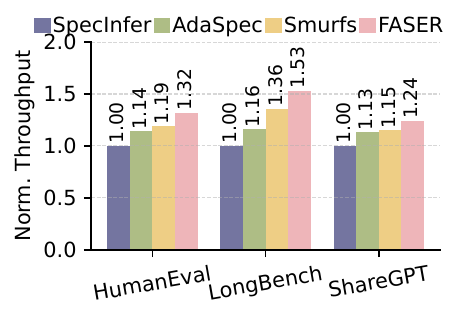}
        \vspace{-1.8em}
        \caption{Qwen3}
        \label{fig:e2e_throughput_Qwen3}
    \end{subfigure}
    \hfill
    \begin{subfigure}[t]{0.48\linewidth}
        \centering
        \includegraphics[width=\linewidth]{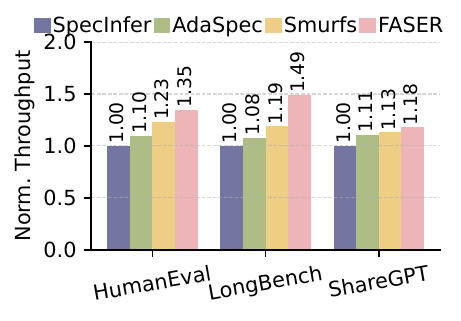}
        \vspace{-1.8em}
        \caption{Llama3}
        \label{fig:e2e_throughput_Llama3}
    \end{subfigure}
    \vspace{-1.2em}
    \caption{Throughput performance of \sys.}
    \vspace{-1.5em}
    \label{fig:e2e_throughput}
\end{figure}

\subsubsection{Throughput.}
\label{sec:throughput}
Fig.~\ref{fig:e2e_throughput} shows the throughput (output tokens per second) of \sys. \sys consistently achieves the highest throughput across all models and datasets. For Qwen3 and Llama3, \sys improves throughput by up to 1.53$\times$ and 1.49$\times$, respectively.
These throughput gains stem from \sys's ability to sustain productive concurrency under dynamic workloads. By finely overlapping generation with verification and reducing unnecessary verification work through early exit, \sys keeps the GPU more continuously occupied with useful work, rather than stalled at stage boundaries or wasted on unnecessary verification. As a result, \sys converts available GPU cycles into more completed output tokens over time.
The baselines achieve lower throughput because their optimizations do not translate into equally effective capacity gains under concurrent serving. SpecInfer and Smurfs still leave verification as a major bottleneck, which limits how efficiently the system can retire tokens at high load. AdaSpec reduces some speculative inefficiency, but the lack of draft-target overlap constrains end-to-end concurrency. In contrast, \sys improves both stage overlap and verification efficiency, allowing it to sustain higher token throughput across a wide range of workloads.

\subsubsection{Runtime Behaviors.}
\label{sec:runtime_behavior}
We further analyze the runtime behaviors of \sys to understand how it achieves the latency reduction and throughput improvement.

\begin{figure}[t]
    \centering
    \begin{subfigure}[t]{0.48\linewidth}
        \centering
        \includegraphics[width=\linewidth]{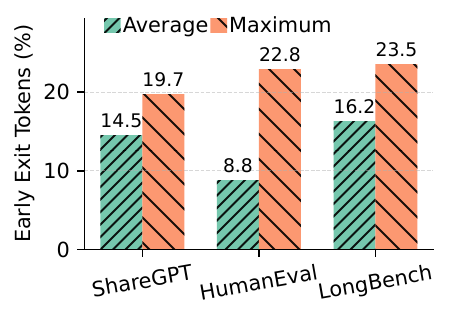}
        \vspace{-1.8em}
        \caption{Qwen3}
        \label{fig:early_exit_tokens_Qwen3}
    \end{subfigure}
    \hfill
    \begin{subfigure}[t]{0.48\linewidth}
        \centering
        \includegraphics[width=\linewidth]{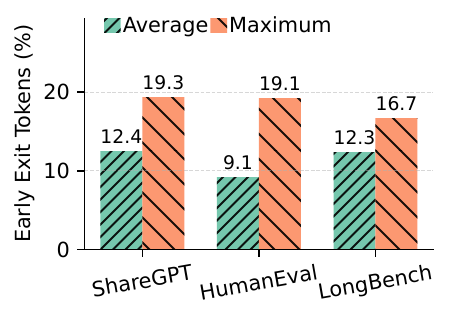}
        \vspace{-1.8em}
        \caption{Llama3}
        \label{fig:early_exit_tokens_Llama3}
    \end{subfigure}
    \vspace{-1.2em}
    \caption{Early exit tokens distribution of \sys.}
    \vspace{-1.5em}
    \label{fig:early_exit_tokens}
\end{figure}

\textbf{Early-Exit Tokens.}
As shown in Fig.~\ref{fig:early_exit_tokens}, token-wise early exiting prunes a fraction of low-quality draft tokens at different layers, which is a key contributor to the strong latency performance of \sys. For example, on HumanEval, Qwen3 early-exits 8.8\% of draft tokens on average, with a maximum of 22.8\%. For Llama3 on the same dataset, \sys early-exits 9.1\% of draft tokens on average, with a maximum of 19.1\%. The early-exit ratio also varies across datasets. On ShareGPT, for instance, the average early-exit ratio is 14.5\% for Qwen3 and 12.4\% for Llama3. Despite these lower pruning rates, \sys still achieves strong end-to-end performance through its joint optimization of early exit, dynamic drafting, and cross-stage overlap, which together reduce the critical-path overhead even when fewer tokens are pruned. Overall, these results show that the early-exit mechanism can effectively identify and remove low-quality draft tokens, thereby improving the efficiency of \sys.

\begin{figure}[t]
    \centering
    \begin{subfigure}[t]{0.48\linewidth}
        \centering
        \includegraphics[width=\linewidth]{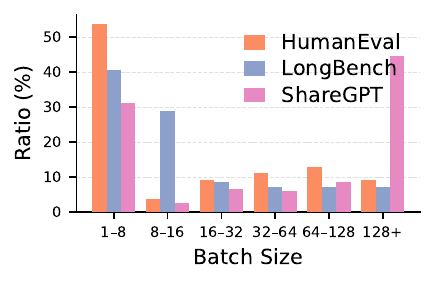}
        \vspace{-1.8em}
        \caption{Various batch sizes}
        \label{fig:batch_size}
    \end{subfigure}
    \hfill
    \begin{subfigure}[t]{0.48\linewidth}
        \centering
        \includegraphics[width=\linewidth]{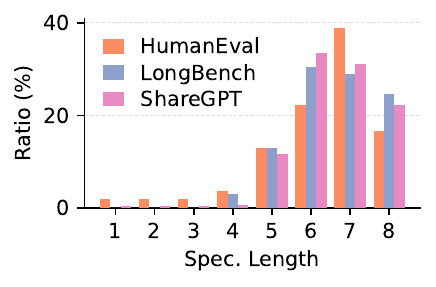}
        \vspace{-1.8em}
        \caption{Various speculative lengths}
        \label{fig:spec_length}
    \end{subfigure}
    \vspace{-1.2em}
    \caption{Batch size and speculative length distribution of \sys of different datasets with Qwen3 model pair.}
    \vspace{-1.2em}
    \label{fig:runtime_distribution}
\end{figure}

\textbf{Distribution of batch size and token length.} 
Fig.~\ref{fig:runtime_distribution} shows the distributions of batch size and speculative length across different datasets for the Qwen3 model pair. It can be observed that highly dynamic workloads lead to markedly different batch-size distributions across datasets. For example, when serving ShareGPT, over 40\% of batches are larger than 128, while about 30\% are smaller than 8. This high variability in batch size allows \sys to achieve higher performance than baselines. For speculative length, the early-exit and fine-grained overlap mechanisms enable \sys to select longer speculative lengths with little additional overhead, with values ranging from 5 to 8 for most datasets.

\subsection{Detailed Analysis}
\label{sec:microscope_analysis}

\begin{figure}[t]
    \centering
    \begin{subfigure}[t]{0.48\linewidth}
        \centering
        \includegraphics[width=\linewidth]{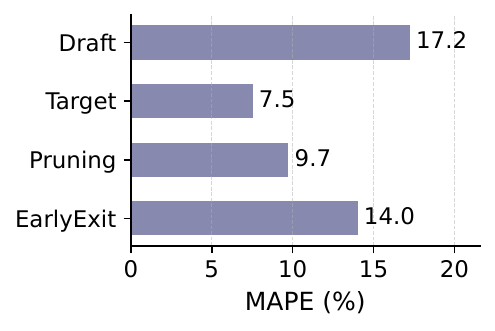}
        \vspace{-1.8em}
        \caption{Qwen3}
        \label{fig:profile_accuracy_Qwen3}
    \end{subfigure}
    \hfill
    \begin{subfigure}[t]{0.48\linewidth}
        \centering
        \includegraphics[width=\linewidth]{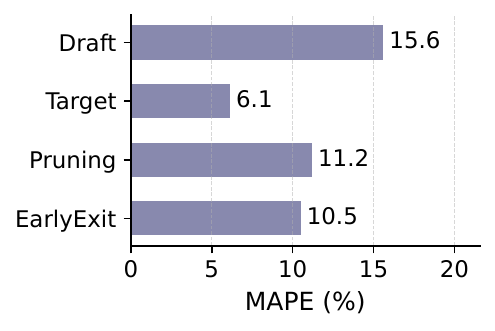}
        \vspace{-1.8em}
        \caption{Llama3}
        \label{fig:profile_accuracy_Llama3}
    \end{subfigure}
    \vspace{-1.2em}
    \caption{Offline profiling accuracy of different profilers.}
    \vspace{-1.5em}
    \label{fig:profile_accuracy}
\end{figure}

\subsubsection{Accuracy of Offline Profiling}
\label{sec:accuracy_offline_profiling}
Accurate latency profiling is critical to \sys, as it directly guides the selection of speculative length, and token-wise exit depth during inference. We evaluate the fitted latency models over batch sizes \{4, 8, \ldots, 256\}, token lengths \{2, 4, 6, 8, 10\}, and SM allocations \{10\%, 20\%, \ldots, 100\%\}. For each configuration, we reserve 20\% of the samples as a held-out set and report the mean absolute percentage error (MAPE) for both draft and target latency.
Fig.~\ref{fig:profile_accuracy} shows the MAPE results for different profiling models across the two model families and three datasets. We find that the MAPE stays below 18\% for most cases, indicating that the profiling method remains accurate and robust across different configurations. The largest error arises for both Qwen3 and Llama3 in draft latency estimation, where the MAPE reaches 17.2\% and 15.6\%, respectively. This is because draft latency is more sensitive to batch size and token length, which can lead to higher variance in measurements and fitting. By contrast, target latency is more stable across configurations, resulting in lower MAPE values of 7.5\% for Qwen3 and 6.1\% for Llama3. Overall, the profiling error remains bounded and is sufficiently small to support the runtime decisions made by \sys.

\begin{figure}[t]
    \centering
    \begin{subfigure}[t]{0.48\linewidth}
        \centering
        \includegraphics[width=\linewidth]{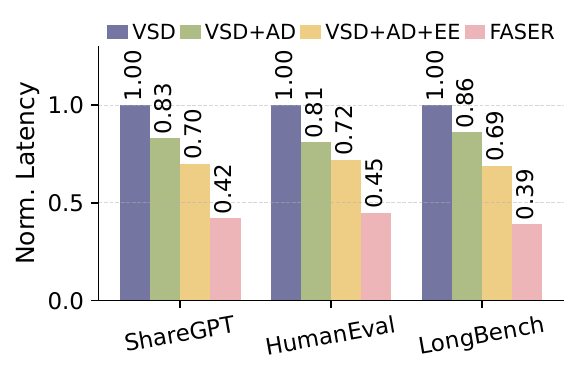}
        \vspace{-1.8em}
        \caption{Norm. Latency}
        \label{fig:each_comp_norm_latency}
    \end{subfigure}
    \hfill
    \begin{subfigure}[t]{0.48\linewidth}
        \centering
        \includegraphics[width=\linewidth]{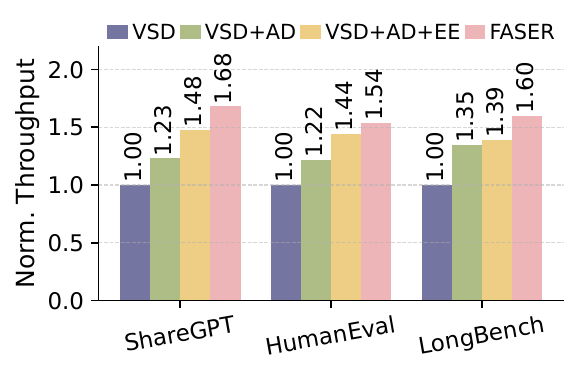}
        \vspace{-1.8em}
        \caption{Norm. throughput}
        \label{fig:each_comp_norm_throughput}
    \end{subfigure}
    \vspace{-1.2em}
    \caption{Effectiveness of each component in \sys with Qwen3 model pair.}
    \vspace{-1.8em}
    \label{fig:each_comp}
\end{figure}

\subsubsection{Effectiveness of Each Component}
\label{sec:effectiveness_of_each_component}
To quantify the contribution of each component in \sys, we conduct an ablation study using the Qwen3 model pair. We measure the performance of \sys by incrementally adding components to vanilla speculative decoding with a fixed speculative length of 4 (VSD). Specifically, VSD+AD augments VSD with \textit{Adaptive Drafter} (AD), while VSD+AD+EE further adds \textit{Token-wise Early Exiter} (EE). \sys incorporates all optimizations across both the draft and target stages. The results are shown in Fig.~\ref{fig:each_comp}.
We find that both AD and EE contribute substantially to the performance gains of \sys. AD reduces latency by up to 19\% and improves throughput by up to 1.35$\times$, while EE reduces latency by 26\% and increases throughput by 1.22$\times$. This difference suggests that EE plays a larger role in reducing latency, as it directly cuts the expensive target-side verification overhead. By contrast, AD contributes more to throughput improvement, since it improves draft-side efficiency and increases the effective benefit of speculation. When combined with \textit{Pipeline Overlapper}, the three components reduce total latency by 61\% and improve throughput by 1.60$\times$, demonstrating the strong synergy of fine-grained overlapping.

\vspace{-.3em}
\subsection{Generalization of \sys}
\label{sec:generalization}
To evaluate the generality of \sys, we extend it to three representative settings: self-speculative decoding (SSD) with Medusa~\cite{cai2024medusa} and EAGLE~\cite{li2024eagle}, and MoE-based models. These results demonstrate that \sys is not restricted to a specific SD framework or model architecture.

\begin{figure}[t]
    \centering
    \begin{subfigure}[t]{0.48\linewidth}
        \centering
        \includegraphics[width=\linewidth]{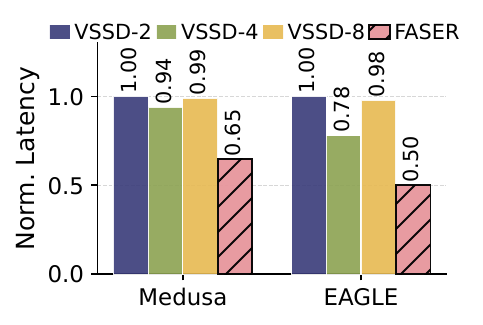}
        \vspace{-1.8em}
        \caption{Norm. Latency}
        \label{fig:self_decoding_latency}
    \end{subfigure}
    \hfill
    \begin{subfigure}[t]{0.48\linewidth}
        \centering
        \includegraphics[width=\linewidth]{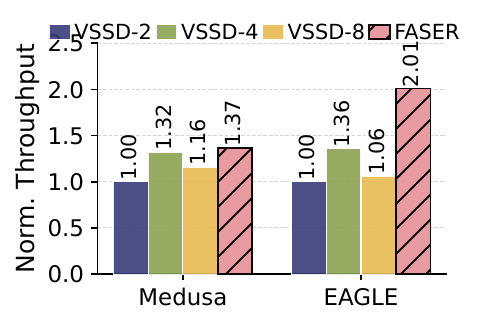}
        \vspace{-1.8em}
        \caption{Norm. throughput}
        \label{fig:self_decoding_throughput}
    \end{subfigure}
    \vspace{-1em}
    \caption{Adaption performance of \sys to self-speculative decoding.}
    \vspace{-1.em}
    \label{fig:adaption_self_decoding}
\end{figure}

\subsubsection{Adaptation to SSD}
\label{sec:adaption_to_self_sd}
SSD is a variant of SD in which both draft and target are derived from the same backbone model. EAGLE performs self-drafting using intermediate-layer features, while Medusa uses auxiliary decoding heads to generate draft tokens and relies on the full model for verification. We adapt \sys to both frameworks by replacing the draft stage in \sys with their respective self-drafting mechanisms, while preserving others as \sys.
We evaluate \sys on top of these SSD frameworks using Qwen3-32B on the ShareGPT dataset with one H100 GPU. We compare \sys against the original implementations of EAGLE and Medusa, as well as vanilla self-speculative decoding (VSSD) baselines with fixed speculative lengths.

For Medusa, \sys achieves up to 35\% lower latency and 1.37$\times$ higher throughput than the original implementation, while VSSD-4 and VSSD-8 improve throughput by 1.32$\times$ and 1.16$\times$ and reduce latency by 6\% and 1\%, respectively. For EAGLE, \sys achieves up to 50\% lower latency and 2.01$\times$ higher throughput than the original implementation, while VSSD-4 and VSSD-8 improve throughput by 1.47$\times$ and 1.89$\times$ and reduce latency by 28\% and 48\%, respectively.
These gains arise because \sys improves the efficiency of self-drafting while reducing low-benefit speculative work during verification. As a result, \sys consistently outperforms both the original SSD implementations and the fixed-length VSSD baselines. The improvements observed for both EAGLE and Medusa suggest that the benefits of \sys generalize across SSD frameworks with different self-drafting mechanisms.

\begin{table}[t]
\centering
\caption{Performance of \sys on MoE models using ShareGPT with 2 H100 GPU device (TP=2).}
\label{tab:adapt_moe}
\vspace{-1em}
\renewcommand{\arraystretch}{0.9}
\small
\setlength{\tabcolsep}{4pt}
\resizebox{\linewidth}{!}{
\begin{tabular}{>{\centering\arraybackslash}p{2.8cm} c c c}
\toprule
\textbf{Model Pair} & \textbf{System} & \textbf{Latency} & \textbf{Throughput} \\
\midrule
\multirow{2}{=}{Qwen2-0.5B/Qwen2-57B-A14B~\cite{yang2024qwen2technicalreport}}
& Smurfs & 1.0 & 1.0 \\
& \sys & \textbf{0.84} & \textbf{1.38} \\
\bottomrule
\end{tabular}}
\end{table}

\subsubsection{Adaptation to MoE models.}
To evaluate the generalization of \sys to MoE models, we adapt \sys to a MoE pair based on the Qwen2 architecture, where the draft model is Qwen2-0.5B and the target model is Qwen2-57B-A14. We evaluate the latency and throughput performance of \sys against Smurfs using the ShareGPT dataset on 2 H100 GPU. The latency and throughput are normalized to the performance of Smurfs.
As shown in Table~\ref{tab:adapt_moe}, \sys achieves a 16\% latency reduction and a 1.38$\times$ throughput improvement over Smurfs on this MoE model pair. These results demonstrate that \sys can effectively adapt to the unique characteristics of MoE models, such as their dynamic expert routing and variable computation patterns, while still delivering strong performance benefits. The improvements are consistent with those observed on dense models, indicating that the core principles of \sys generalize well to different model architectures.

\begin{figure}[t]
    \centering
    \begin{subfigure}[t]{0.48\linewidth}
        \centering
        \includegraphics[width=\linewidth]{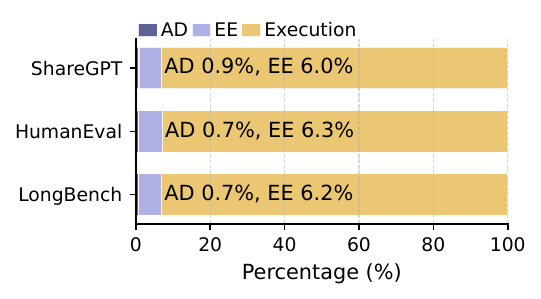}
        \vspace{-1.8em}
        \caption{Qwen3}
        \label{fig:qwen3_phase_percentage}
    \end{subfigure}
    \hfill
    \begin{subfigure}[t]{0.48\linewidth}
        \centering
        \includegraphics[width=\linewidth]{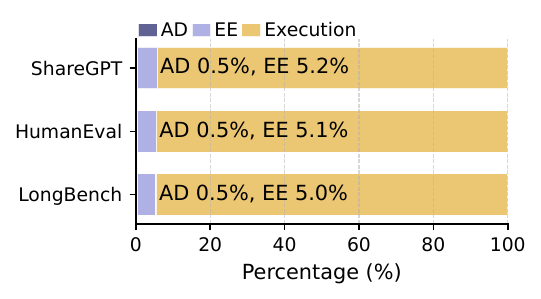}
        \vspace{-1.8em}
        \caption{Llama3}
        \label{fig:llama3_phase_percentage}
    \end{subfigure}
    \vspace{-1.2em}
    \caption{System overhead of \sys.}
    \vspace{-1em}
    \label{fig:system_overhead}
\end{figure}

\vspace{-.3em}
\subsection{System Overhead}
\label{sec:system_overhead}
We evaluate the runtime overhead introduced by \textit{Adaptive Drafter} (AD) and \textit{Token-wise Early Exiter} (EE) in \sys. AD incurs overhead when selecting the speculative length, while EE adds cost to determine whether a draft token should exit early. We quantify this overhead as the fraction of total inference time spent in each component.
Fig.~\ref{fig:system_overhead} shows that the overhead remains small across all models and datasets. AD accounts for only 0.5\%$\sim$0.9\% of total inference time, while EE accounts for 5.0\%$\sim$6.3\%. EE is more expensive because it performs additional token-level checks during verification at each model layer, but its cost remains bounded.
Overall, these results show that the control overhead of \sys is low. The added cost of AD and EE is modest relative to end-to-end execution and is outweighed by the latency and throughput gains of \sys.

%% file: sections/7-related_work.tex
\vspace{-.5em}
\section{Related Work} 
\label{sec:related_work}
\textbf{Improving acceptance.}
Recent work has extensively studied how to improve the acceptance ratio of speculative decoding~\cite{xia2024unlocking,zhang2024beyond,zhang2024draft,zimmer2024mixture}. One line of work focuses on better candidate selection. SpecInfer~\cite{miao2024specinfer}, Minions~\cite{wang2024minions}, and SpecExec~\cite{svirschevski2024specexec} construct and search token trees to select more promising speculative tokens, while BanditSpec~\cite{hou2025banditspec} adaptively chooses draft models or speculation lengths to increase accepted tokens. Another line of work improves acceptance by augmenting the target model itself, as in Medusa~\cite{cai2024medusa} and Amphista~\cite{li2024amphista}, which attach predictive heads to generate speculative tokens. In contrast, \sys does not optimize acceptance alone; instead, it jointly considers acceptance behavior and verification latency under dynamic serving conditions.

\noindent\textbf{Early-exit-based and self-speculative decoding.}
A related line of work applies early exiting to reduce verification overhead in speculative decoding. In self-speculative decoding, methods such as LayerSkip~\cite{elhoushi2024layerskip}, Kangaroo~\cite{liu2024kangaroo}, and related variants~\cite{bae2023fast,zhang2024draft,zhang2024draft2,liu2024speculative,xia2024swift} reuse early layers of the target model as an internal draft model and use later layers for verification, thereby avoiding a separate drafter. Beyond self-speculation, HiSpec~\cite{kumar2025hispec} introduces trained early-exit models as intermediate verifiers to discard low-quality tokens before full verification. SpecEE~\cite{xu2025specee} further accelerates speculative early exiting with lightweight predictors and system-level scheduling, and is orthogonal to \sys. In contrast, \sys performs dynamic system-level token-wise early exiting during verification under the current serving condition, without requiring model-specific early-exit structures or specialized early-exit training.

\noindent\textbf{Pipeline overlapping.}
Recent research~\cite{butler2024pipeinfer,liu2025pearl,wang2025towards,mcdanel2025pipespec,yin2025specpipe,shen2026specbranch} increasingly overlaps drafting and verification to reduce mutual waiting. PipeInfer~\cite{butler2024pipeinfer} uses asynchronous pipelining, but is mainly designed for single-request distributed inference. PEARL~\cite{liu2025pearl} overlaps the two stages through parallel verification and drafting. Smurfs~\cite{wang2025towards} pipelines SSM speculation and LLM verification across batches, but mainly hides SSM-side latency and does not explicitly partition GPU resources between the draft and target models. PipeSpec~\cite{mcdanel2025pipespec} and SpecPipe~\cite{yin2025specpipe} extend this direction to hierarchical or pipeline-parallel settings. SpecBranch~\cite{shen2026specbranch} further reduces serialized dependence between drafting and verification through branch parallelism. Overall, 
these works show the benefit of draft-target overlap, but they largely focus on coarse-grained pipelining and do not consider dynamic system conditions or fine-grained pipeline overlapping, which is the focus of \sys.

%% file: sections/8-conclusion.tex
\section{Conclusion}
\label{sec:conclusion}
This paper introduces \sys, a system that replaces rigid, coarse-grained speculative decoding with fine-grained management to handle dynamic LLM workloads. By combining request-level adaptive drafting, early exiting of rejected tokens, and hardware-aware stage overlapping via spatial multiplexing, \sys effectively eliminates serialized bottlenecks and computational waste. 


\section{Acknowledgments}
The authors thank the members of the HyScale lab at NTU Singapore for their constructive discussions and feedback on this work. This project is supported by the Ministry of Education, Singapore, under its Academic Research Funds Tier 1 RG110/25.